\pdfoutput=1
\documentclass[letter]{aa}  

\usepackage{graphicx}
\usepackage{txfonts}
\newcommand{\BE}{\begin{equation}}
\newcommand{\EE}{\end{equation}}
\newcommand{\BA}{\begin{eqnarray}}
\newcommand{\EA}{\end{eqnarray}}

\renewcommand{\vec}[1]{{\boldsymbol #1}}
\newcommand{\deriv}[2]{\frac{{\mathrm d} #1}{{\mathrm d} #2}}
\newcommand{\rmd}{ {\ \mathrm d} }

\newcommand{\surf}{ {\mathcal S} }

\newcommand{\bb}{\vec B}

\newcommand{\xx}{ \vec x}
\newcommand{\uu}{ \vec u}
\newcommand{\vv}{ \vec v}
\newcommand{\bn}{B_{n}}

\newcommand{\gth}{G_{\theta}}
\newcommand{\gph}{G_{\Phi}}

\newcommand{\eq}[1]{Equation~(\ref{eq:#1})} 
\newcommand{\eqs}[2]{Equations~(\ref{eq:#1}) and (\ref{eq:#2})} 
 
\newcommand{\sect}[1]{Section~\ref{sec:#1}}

\newcommand{\fig}[1]{Figure~\ref{fig:#1}}

\newcommand{\eg}{\textit{e.g.}, }
\newcommand{\ie}{\textit{i.e.}, }

\begin{document} 

    \title{First observational application of a connectivity--based \\ helicity flux density}

   \subtitle{}

   \author{K. Dalmasse
          \inst{1}
          \and
          E. Pariat\inst{1}
          \and
          G. Valori\inst{1}
          \and
          P. D\'emoulin\inst{1}
                    \and
          L. M. Green\inst{2}
          }

   \institute{LESIA, Observatoire de Paris, CNRS, UPMC, 
   		Universit\'e Paris-Diderot, 92195 Meudon, France \\
                   \email{kevin.dalmasse@obspm.fr}
         \and
                   Mullard Space Science Laboratory, Univ. College London, U.K.
             }

   \date{}

 
  \abstract
    {
Measuring the magnetic helicity distribution in the solar corona 
can help in understanding the trigger of solar eruptive events because magnetic 
helicity is believed to play a key role in solar activity due to its 
conservation property.
    }
    {
A new method for computing the photospheric distribution of the helicity flux 
was recently developed. This method takes into account the magnetic field connectivity 
whereas previous methods were based on photospheric signatures only. 
This novel method maps the true injection of magnetic helicity in active regions. 
We applied this method for the first time to an observed active region, NOAA 11158, 
which was the source of intense flaring activity.
    }
    {
We used high--resolution vector magnetograms from the SDO/HMI
instrument to compute the photospheric flux transport velocities and to 
perform a nonlinear force--free magnetic field extrapolation. We 
determined and compared the magnetic helicity flux distribution using 
a purely photospheric as well as a connectivity--based method.
    }
  {
While the new connectivity--based method confirms the mixed pattern 
of the helicity flux in NOAA 11158, it also reveals a different, and more 
correct, distribution of the helicity injection. This distribution can be important 
for explaining the likelihood of an eruption from the active region.
  }
    {
The connectivity–based approach is a robust method for computing the magnetic helicity 
flux, which can be used to study the link between magnetic helicity and eruptivity of observed 
active regions.
   }
   

   \titlerunning{First observational application of a connectivity--based helicity flux density}
   \authorrunning{K. Dalmasse et al}

   \maketitle
   
%
\section{Introduction}  \label{sec:S-Introduction}

Magnetic helicity globally characterizes the geometrical properties of the magnetic field 
in a volume, generalizing more local properties such as twist and shear \citep[\eg][]{Moffatt69}. 
Because of its conservation property \citep[see][]{Berger84}, magnetic helicity is believed to 
play a key role in solar eruptivity. In particular, it has been argued that magnetic helicity 
accumulation and/or annihilation can be involved in the generation and dynamics of solar 
flares and coronal mass ejections \citep[CMEs; \eg][]{Low97,Kusano02}.

Observationally, evidence of magnetic helicity accumulation and annihilation can be obtained 
from studying the photospheric distribution of the magnetic helicity flux in active regions (ARs) 
prior to flaring activity \citep[\eg][]{Moon02,Chae04,Labonte07}. Magnetic helicity accumulation 
is straightforward when the helicity flux is uniformly distributed in sign. In a given magnetic system, 
a sequential injection of helicity of different signs would reduce the accumulated helicity. 
Magnetic helicity annihilation, by contrast, necessitates a distribution with opposite helicity 
fluxes since the annihilation of helicity requires the reconnection of magnetic domains of 
opposite helicity. \cite{Linton01} showed that magnetic reconnection can release more energy 
when it occurs between systems of opposite helicity, and \cite{Kusano02,Kusano04} developed 
a model of solar flares based on the interaction of magnetic structures of opposite helicity.

Observational studies were therefore carried out to detect ARs with opposite helicity 
fluxes \citep[\eg][]{Chandra10,Romano11,Jing12,Vemareddy12d}. The helicity flux 
distribution in ARs is measured by computing a helicity flux density proxy, since helicity is 
not a local quantity \citep[see][]{Pariat05}. The proxies usually employed only require the 
photospheric evolution of AR vector magnetograms. These proxies do not allow a direct 
interpretation of the helicity flux density distribution when opposite signals of helicity flux are 
present, and can even introduce spurious, \ie fake, signals \citep[see][]{Pariat05,Pariat07c}.

\cite{Pariat05} proposed a new proxy of the helicity flux density that takes into account the 
magnetic field connectivity, and \cite{Dalmasse13} developed a method to compute such 
a proxy in practice. Based on analytical case studies and numerical simulations, they showed 
that this method can reliably and accurately determine the injection of helicity, and can reveal 
the real mixed signals of the helicity flux. Therefore, this method should be used to map the 
helicity flux in observational studies.

As an example, recent observational studies of the photospheric distribution of the 
magnetic helicity flux in NOAA 11158 found mixed signals of the helicity flux in the AR 
as reported by \cite{Jing12} and \cite{Vemareddy12d}. These authors argued that the pattern 
can be associated to the injection of opposite helicity and that magnetic helicity annihilation 
can be responsible for triggering some major flares observed in the AR. However, they 
employed the purely photospheric proxy only instead of a connectivity--based helicity flux 
density method to compute the photospheric distribution of the helicity flux in the AR. 
We can therefore wonder if these mixed signals are real or spurious. In particular, how 
much are the helicity flux distribution and the signal intensity modified when using 
a connectivity--based helicity flux density proxy? Hence, can the obtained helicity flux 
distribution give hints of the action of magnetic helicity annihilation in the activity of 
NOAA 11158? 

In the present letter, we address these questions by comparing the helicity flux distribution in 
NOAA 11158 computed with the purely photospheric as well as the connectivity--based proxy. 
We present the first observational application of the connectivity--based helicity flux density 
method introduced in \cite{Dalmasse13} and test the robustness of this method.

\section{Observations and analysis} \label{sec:S-Obs-and-method}
\subsection{Method} \label{sec:S-Method}

\cite{Pariat05} demonstrated that the helicity flux can be expressed as the summation of 
the relative rotation rate of all pairs of elementary magnetic flux tubes weighted by their 
magnetic flux. From this definition, they introduced a purely photospheric proxy of helicity 
flux density, $\gth$, such that
    \BE \label{eq:Eq-Gtheta}
      \gth(\xx) = - \frac{\bn}{2\pi} \int_{\surf'} 
                                         \frac{ \left( (\xx-\xx') \times (\uu-\uu') \right) |_{n}}{|\xx-\xx'|^{2}}
                                        \bn' \rmd \surf' \,,
    \EE
where $S'$ is the photospheric--surface where the helicity flux is computed, 
$\xx$ and $\xx'$ are the photospheric magnetic footpoints of elementary magnetic 
flux tubes, $\bb$ and $\bb'$ are their associated magnetic fields (index $n$ or $t$ 
denotes the normal or transverse component of a vector), and $\uu$ and $\uu'$ are 
their respective flux transport velocities defined by \citep[\eg][]{Demoulin03}:
    \BE \label{eq:Eq-FTV}
      \uu =  \vv_{t} - \frac{v_{n}}{\bn} \bb_{t} \,,
    \EE
where $\vv$ is the photospheric plasma velocity.

\eqs{Eq-Gtheta}{Eq-FTV} show that at a given time, $t$, the photospheric distribution 
of helicity flux, $\gth$, can be computed from a timeseries of vector magnetograms. 
However, as pointed out by \cite{Pariat05}, a helicity flux density is only meaningful when 
considering a whole elementary magnetic flux tube/field line. Such an extension implies taking 
into account the magnetic field connectivity, which is not included in $\gth$. To solve this 
problem, \cite{Pariat05} defined a connectivity--based helicity flux density proxy, $\gph$, 
such that
      \BE \label{eq:Eq-Gphi}
	\gph(\xx_{c_{\pm}}) = \frac{1}{2} \left( \gth(\xx_{c_{\pm}}) + 
	         \left| \frac{\bn(\xx_{c_{\pm}})}{\bn(\xx_{c_{\mp}})}\right| \gth(\xx_{c_{\mp}}) \right) \,,
      \EE
where {\it c} is a closed elementary magnetic flux tube, \ie a field line, anchored in the 
photosphere at the magnetic footpoints $\xx_{c_\pm}$.
      
\cite{Dalmasse13} introduced a method for computing \eq{Eq-Gphi} based on magnetic 
field line integration, which requires the knowledge of the magnetic field in the volume 
above the region of interest. Because magnetic field measurements are mostly realized 
at the photospheric level, we performed magnetic field extrapolations to obtain the coronal 
magnetic field and to compute the photospheric distribution of the magnetic helicity 
from \eq{Eq-Gphi}.

Additionally, we computed the true density of the helicity flux for each elementary magnetic 
flux tube {\it c}, \ie the helicity flux per unit magnetic flux, $\rmd h_{\Phi} / \rmd t$, defined by 
\citep[see][]{Pariat05}
      \BE \label{eq:Eq-dhphi}
         \deriv{h_{\Phi}}{t}\bigg|_{c} =  \frac{\gth(\xx_{c_{+}})}{|\bn(\xx_{c_{+}})|} 
                                                             + \frac{\gth(\xx_{c_{-}})}{|\bn(\xx_{c_{-}})|} \,.
      \EE
    
By definition from \eqs{Eq-Gphi}{Eq-dhphi}, $\gph$ and $\rmd h_{\Phi} / \rmd t$ 
are defined only for closed magnetic field lines. For open magnetic field lines, we set 
$\gph=\gth$ and did not compute $\rmd h_{\Phi} / \rmd t$. Finally, closed magnetic field 
lines with $\bn(\xx_{c_{\pm}})$ lower than $10$ gauss at one footpoint were treated 
as open field lines to avoid problems related to the presence of bald patches that would 
result in very high/infinite values of the helicity flux density, and that would prevent the 
conservation of the total helicity flux.

  \begin{figure} [h]   
   \centerline{\includegraphics[width=0.42\textwidth,clip=]{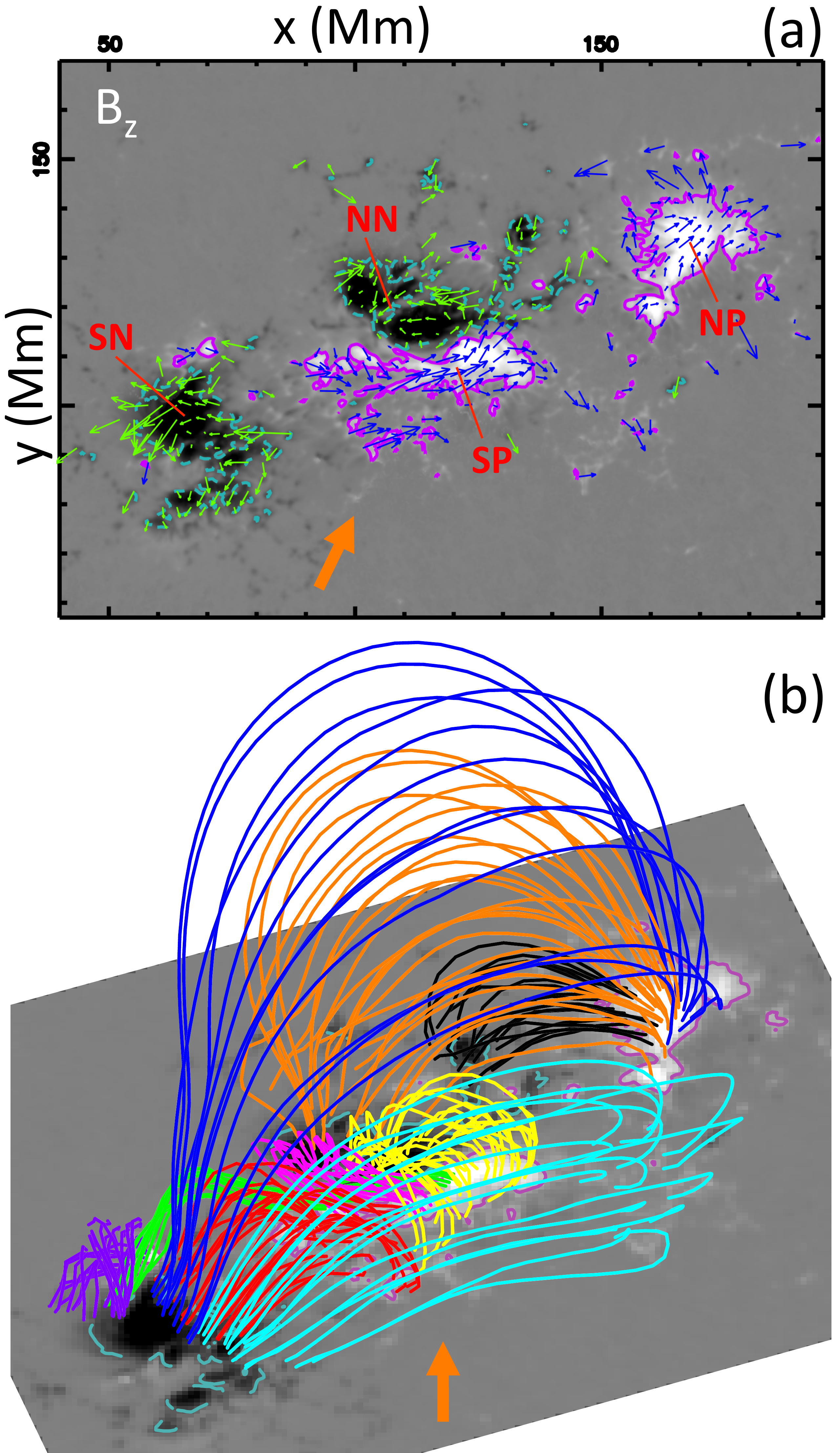}
              }
              \caption{Solar active region NOAA 11158 at $\sim$ 06:28 UT on 2011 February 14. (a) Photospheric vertical magnetic field ($B_z$) in grayscale overplotted with the flux transport velocity field (blue/green arrows) and the polarity labels (red). (b) 3D views of the NLFFF magnetic field extrapolation with selected magnetic field lines. Each field line color corresponds to a different quasi--connectivity domain (separated by QSLs; see \sect{S-Conclusion-Discussion}). The orange arrow shows the viewing angle relatively to panel (a). The magnetic field values are saturated at $\pm 1500$ gauss. Solid pink and dashed cyan lines are $\pm 500$ gauss magnetic field isocontours, respectively.
                      }
   \label{fig:Fig-Bz-FTV-FLs}
   \end{figure} 

\subsection{Data}      \label{sec:S-Data}

NOAA 11158 appeared on 2011 February 10 at the heliographic coordinates S19 E42. 
This AR was the result of strong and fast magnetic flux emergence that resulted in the formation 
of two large--scale bipoles, a northern and a southern one \citep[\eg][]{Schrijver11}. 
Its emergence was associated to several C-, M-, and X-class flares and CMEs during 2011 
February 10--20 \citep[\eg][]{Sun12}.

We used the 12 minute cadence and high--resolution vector magnetograms from the SDO/HMI 
instrument at 06:22 and 06:34 UT on 2011 February 14 prior to a C-class flare observed at 
06:51 UT. For both vector magnetograms, the $180^{\circ}$ ambiguity was removed using the 
method of \cite{Metcalf06}. These vector magnetograms were used to compute the photospheric 
flux transport velocity field using the differential affine velocity estimator for vector magnetograms 
\citep[DAVE4VM;][]{Schuck08} with a window size of 19 pixels \citep[as in][]{Liu13}.

We assumed that the computed flux transport velocity field is equivalent to an instantaneous flux 
transport velocity at $\sim$ 06:28 UT. We then constructed the associated vector magnetogram by 
averaging the magnetic data measurements taken at 06:22 and 06:34 UT (\fig{Fig-Bz-FTV-FLs}a).

We performed a nonlinear force--free magnetic field (NLFFF) extrapolation to obtain the coronal 
magnetic field from the vector magnetogram at $\sim$ 06:28 UT using the magnetofrictional 
relaxation method developed in \cite{Valori05,Valori10}. The data were first rebinned to $1''$ 
per pixel and preprocessed toward the force--free condition using the method of \cite{Fuhrmann07}. 
The extrapolation domain covers $\sim 208 \times 202 \times 145 \ \textrm{Mm}^{3}$. A set of 
selected magnetic field lines is represented in \fig{Fig-Bz-FTV-FLs}b.

\section{Results} \label{sec:S-Results}

In the following, the positive and negative polarities of the northern or southern bipole are 
referred to as NP and NN, or SP and SN, respectively (see \fig{Fig-Bz-FTV-FLs}a).

\begin{figure} [t] 
   \centerline{\includegraphics[width=0.42\textwidth,clip=]{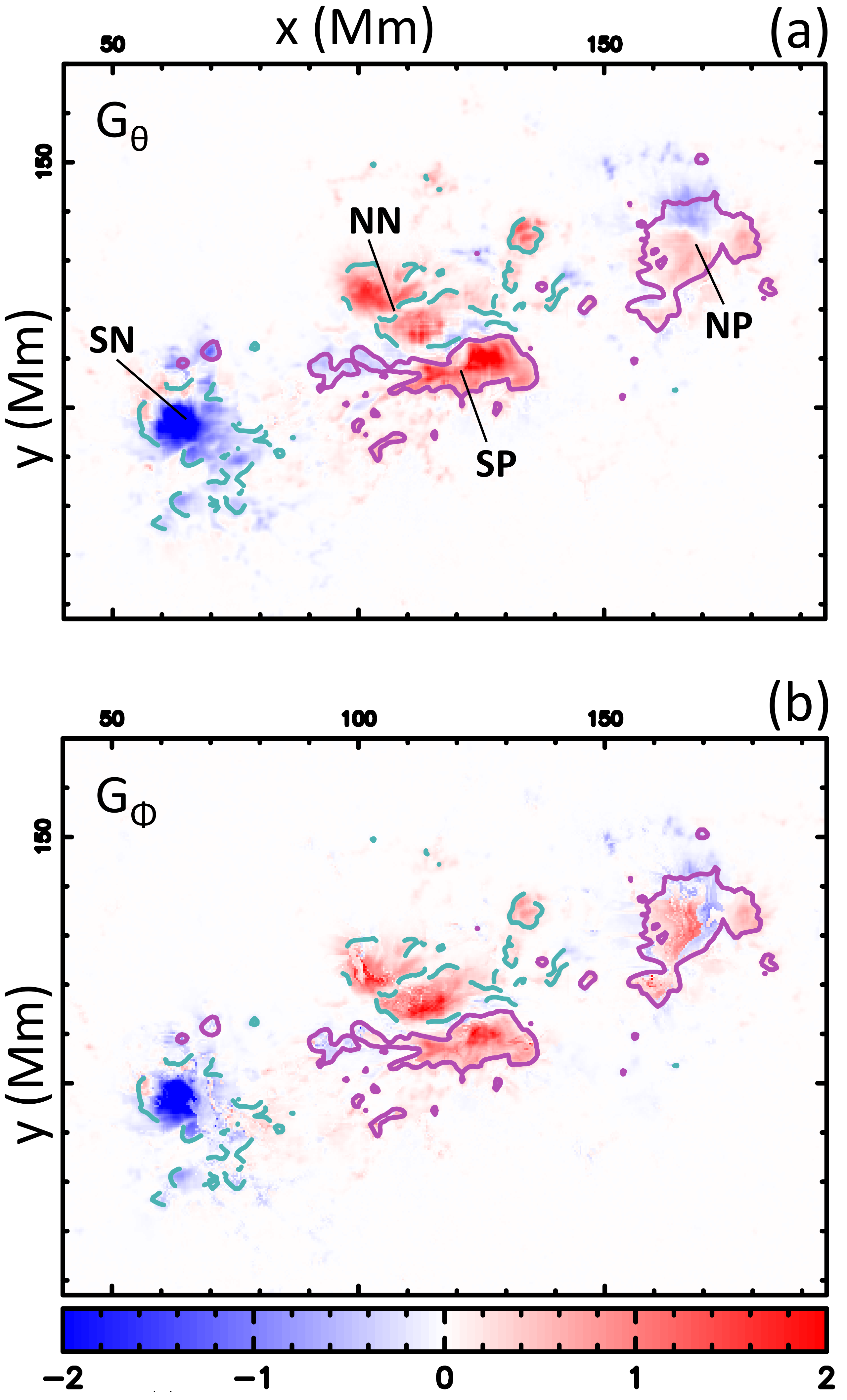}
              }
              \caption{Solar active region NOAA 11158 at $\sim$ 06:28 UT on 2011 February 14. (a,b) Helicity flux density distributions in units of $10^7 \ \textrm{Wb}^2 \ \textrm{m}^{-2} \ \textrm{s}^{-1}$ with the same color scale: (a) $\gth$ and (b) $\gph$ maps. All photospheric maps are overplotted with $\pm 500$ gauss isocontours of $B_z$.
                      }
   \label{fig:Fig-SurfaceHflux}
   \end{figure}

The full photospheric $xy$-domain of the NLFFF extrapolation was considered to derive 
the helicity flux density maps displayed in \fig{Fig-SurfaceHflux}. 
In this domain, the closed magnetic field --- for which $\gph$ has been 
computed --- encloses $73 \%$ of the total unsigned magnetic flux. The remaining magnetic flux 
corresponds to open--like magnetic fields (see \sect{S-Method}). These regions, 
where $\gph=\gth$, are mainly located at the eastern and western extremities of the AR.

We compared the total fluxes computed from both $\gth$ and $\gph$ maps displayed in 
\fig{Fig-SurfaceHflux}. The helicity flux of the closed magnetic field and the total helicity 
flux computed with the proxy $\gth$ are $3.7 \times 10^{21} \ \textrm{Wb}^2 \ \textrm{s}^{-1}$ and 
$2.3 \times 10^{21} \ \textrm{Wb}^2 \ \textrm{s}^{-1}$, respectively. The total helicity flux 
is lower because of the strong contribution of negative helicity in the open magnetic field. 
The same fluxes computed with the proxy $\gph$ are 
$3.9 \times 10^{21} \ \textrm{Wb}^2 \ \textrm{s}^{-1}$ and 
$2.5 \times 10^{21} \ \textrm{Wb}^2 \ \textrm{s}^{-1}$. The ratio of these fluxes computed 
with $\gph$ compared with $\gth$ are $1.05$ and $1.08$, respectively. 
Therefore, the global helicity flux between the two proxies agrees very well.

In \fig{Fig-SurfaceHflux}b, the $\gph$ map also displays mixed signals. This implies that there 
are real mixed signs of the helicity flux in the AR, as found in previous studies 
\citep[\eg][]{Jing12,Vemareddy12d}. However, the distribution is different from the results of the 
$\gth$ map (\fig{Fig-SurfaceHflux}a) except in the regions of open magnetic fields. 
In SN, the weak positive signal has almost entirely been replaced by negative 
helicity flux. In NP, $\gph$ presents enhanced positive values on the left side and an intrusion 
of negative values in the central part. Finally, the weak negative signal present between NN and 
SP in $\gth$ ($\approx -2.5 \times 10^6 \ \textrm{Wb}^2 \ \textrm{m}^{-2} \ \textrm{s}^{-1}$) is 
now positive in $\gph$ ($\approx 2.5 \times 10^6 \ \textrm{Wb}^2 \ \textrm{m}^{-2} \ \textrm{s}^{-1}$). 
Otherwise, the distribution of $\gph$ and $\gth$ is similar in NN and SP because field lines are 
linking similar values of $\gth$.

Considering only the closed magnetic field, we summed the positive and then the negative 
helicity flux signals for both $\gth$ and $\gph$. We found that the ratio of these fluxes computed 
with $\gph$ compared with $\gth$ are $0.84$ and $0.57$, respectively. We summed the 
absolute value of these positive and negative fluxes to obtain the total unsigned helicity flux for 
both $\gph$ and $\gth$, and found a ratio equal to $0.76$. Overall, it shows 
that the intensity of the signal in $\gth$ tends to be overestimated.

\begin{figure} [t] 
   \centerline{\includegraphics[width=0.41\textwidth,clip=]{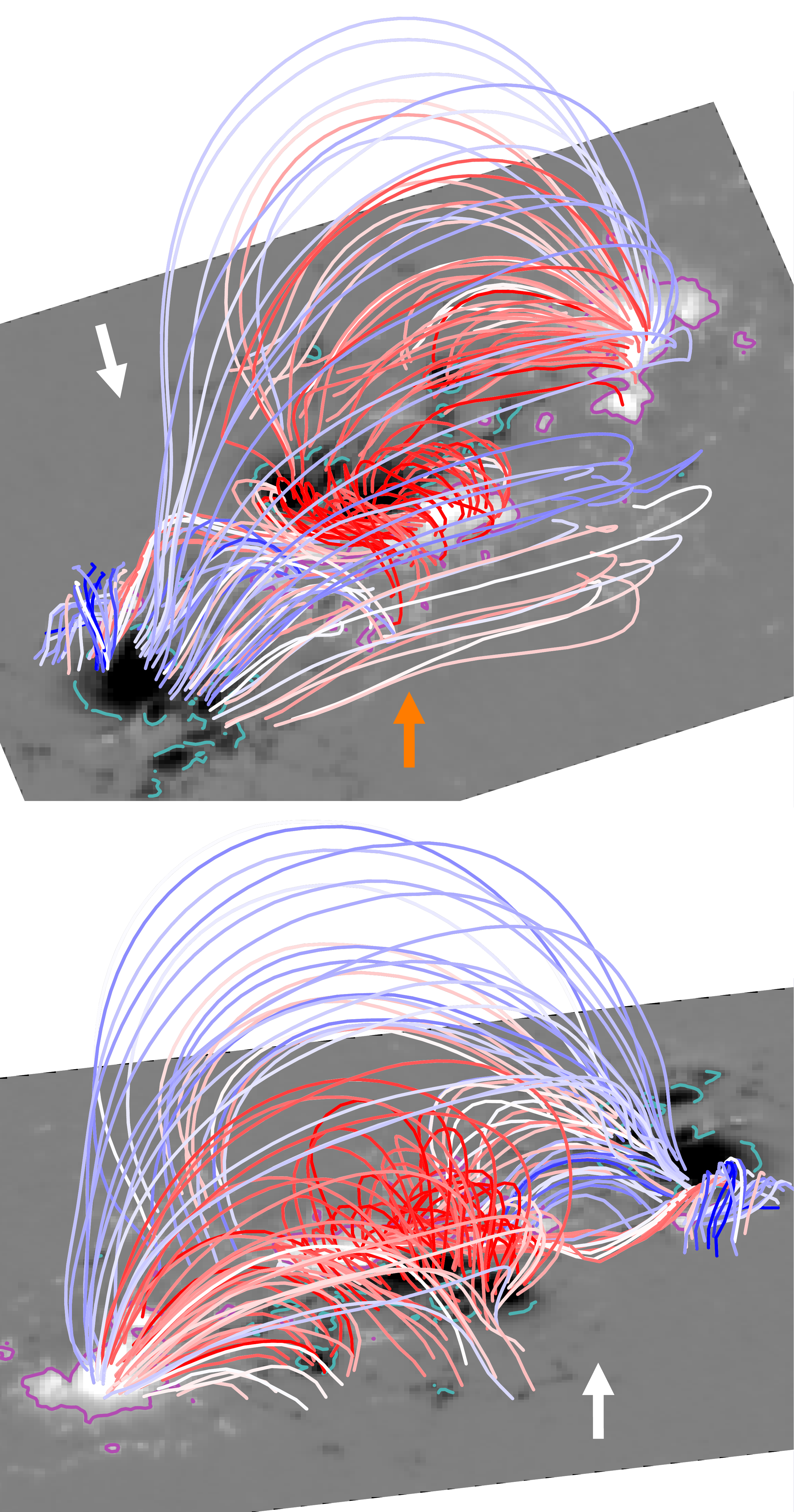}
              }
              \caption{Two 3D views of the NLFFF magnetic field extrapolations with the same selected magnetic field lines as in \fig{Fig-Bz-FTV-FLs}b. Field lines are colored according to their associated helicity flux density value computed from \eq{Eq-dhphi}, red or blue for positive or negative values. The values are saturated at $\pm 1.5 \times 10^8 \ \textrm{Wb} \ \textrm{s}^{-1}$. Arrows show the viewing angles relative to \fig{Fig-Bz-FTV-FLs} (orange) and to the bottom panel (white).
                      }
   \label{fig:Fig-Hfluxdens}
   \end{figure}

In \fig{Fig-Hfluxdens}, we represent the 3D distribution of the helicity flux density per elementary 
magnetic flux tubes/field lines. In this figure, each magnetic field line is colored according to the 
associated true helicity flux density computed from \eq{Eq-dhphi}. The 3D representation of the true 
helicity flux density shows us that the helicity flux density map (\fig{Fig-SurfaceHflux}b) is the result 
of two magnetic structures of strong opposite helicity flux: an inner system with positive 
helicity flux overlaid by an outer system of negative helicity flux.

We also computed the helicity flux distribution using a potential magnetic field extrapolation. 
The results also show mixed helicity flux signals. Compared with the NLFFF case, the differences 
are mostly located in the highly sheared region between SP and NN where the electric currents are 
the strongest. In particular, the intensity of the signal is different, but not its sign. This mutual 
consistency between the application to potential and NLFFF extrapolations, together with the tests 
performed in \cite{Dalmasse13}, provides us with additional confidence that our method for 
computing $\gph$ is robust when applied to observational data.

\section{Conclusions and discussion} \label{sec:S-Conclusion-Discussion}

We presented the first observational application of a connectivity--based helicity flux density 
proxy to NOAA 11158 using the method developed in \cite{Dalmasse13}. We showed that 
the method can be reliably applied to observational data, and should be used to monitor the 
injection of magnetic helicity in ARs.

By computing the photospheric mapping of the helicity flux using this method with an NLFFF 
extrapolation, we provided direct and reliable evidence that real opposite helicity fluxes were 
observed in NOAA 11158 on 2011 February 14. However, the intensity and the distribution 
of the helicity flux are different from what has been found in previous studies 
\citep[][]{Jing12,Vemareddy12d}. The connectivity--based helicity flux density proxy shows 
that (1) the intensity of the fluxes tends to be overestimated with $\gth$, and (2) that two 
magnetic structures with opposite helicity flux are present, one on top of the other.

Although it is not included here, we computed the photospheric mapping of quasi--separatrix 
layers \citep[QSLs; see review by][and references therein]{Demoulin06Review}. These are 
regions of sharp gradients of the magnetic field connectivity and are preferential sites for 
current layers formation, and thus, for magnetic reconnection \citep[\eg][]{Janvier13}. Comparing 
the QSLs mapping with the distribution of helicity flux, we found several locations where the 
interface between the regions of opposite helicity flux coincides with QSLs. 
One such particularly interesting region is located in the northern part of the southern bipole. 
There, we found a flux rope (green field lines in \fig{Fig-Bz-FTV-FLs}b) linking SN to SP, with a 
net positive helicity flux (\fig{Fig-Hfluxdens}). The negative footpoints of this flux rope are below 
surrounding twisted and arcade field lines (purple field lines on the north of SP in 
\fig{Fig-Bz-FTV-FLs}b) which are associated to a strong negative helicity flux (\fig{Fig-Hfluxdens}). 
Hence, the helicity flux pattern could possibly play a role  in the initiation and dynamics of 
the C-class flare observed in this region at 06:51 UT, \ie 20 minutes after our extrapolation 
and helicity injection map. Assuming that these opposite helicity fluxes correspond to a 
transfer of opposite helicity from the convection zone toward the corona, NOAA 11158 
would be a good candidate for studying solar eruptivity related to magnetic helicity annihilation 
\citep[\eg][]{Kusano02,Kusano04}.

Overall, the temporal evolution study of the connectivity--based helicity flux density and of the helicity 
flux density per elementary magnetic flux tube is needed to obtain information on the role of helicity 
injection in the flaring activity observed in ARs, and on the nature of the 
emerging magnetic flux tube(s) creating these ARs.

\begin{acknowledgements} 
This work used the DAVE4VM code written and developed by P. Schuck at the Naval Research Laboratory. 
The research leading to these results has received funding from the European Commission's Seventh 
Framework Programme (FP7/2007-2013) under the grant agreement eHeroes 
(project n$^\circ$~284461, www.eheroes.eu). 
Lucie Green is grateful to the Royal Society for funding through their University Research Fellowships. 
The data used here are courtesy of the NASA/SDO and the HMI science team.
\end{acknowledgements}

\bibliographystyle{aa}
      
\bibliography{PhotInj_NOAA11158_Dalmasse_etal_13}  

\IfFileExists{\jobname.bbl}{} {\typeout{}
\typeout{****************************************************}
\typeout{****************************************************}
\typeout{** Please run "bibtex \jobname" to obtain} \typeout{**
the bibliography and then re-run LaTeX} \typeout{** twice to fix
the references !}
\typeout{****************************************************}
\typeout{****************************************************}
\typeout{}}

\end{document}